\documentclass[mathpazo]{cicp}
\usepackage{hyperref}
\usepackage{graphicx}
\usepackage{tabularx}
\usepackage{amsfonts}
\usepackage{amssymb}
\usepackage{amsmath}
\usepackage{wasysym}
\usepackage{color}
\usepackage{ulem}
\begin{document}
\title{Morphological Similarities between Single-walled\\ Nanotubes and Tubelike Structures of Polymers\\
 with Strong Adsorption Affinity to Nanowires}

\author[T.~Vogel et.~al]{Thomas Vogel\affil{1}\comma\corrauth,
  Tali Mutat\affil{2},
  Joan Adler\affil{2},
  and Michael Bachmann\affil{1}}

\address{%
\affilnum{1}\ Center for Simulational Physics, Department of Physics and Astronomy,\\
The University of Georgia, Athens, GA, 30602, USA\\
\affilnum{2}\ Department of Physics, Technion, Israel Institute of Technology, Haifa, 32000, Israel}

\emails{%
  {\tt thomasvogel@physast.uga.edu} (T.~Vogel),\\
  {\tt talimu@techunix.technion.ac.il} (T.~Mutat),
  {\tt phr76ja@techunix.technion.ac.il} (J.~Adler),\\
  {\tt bachmann@smsyslab.org} (M.~Bachmann)}

\begin{abstract}
  In their tubelike phase, nanowire-adsorbed polymers exhibit strong
  structural similarities to morphologies known from single-walled
  carbon (hexagonal) and boron (triangular) nanotubes.  Since
  boron/boron nitride tubes require some disorder for stability the
  triangular polymer tubes provide a closer analog to the carbon
  tubes.  By means of computer simulations of both two and three
  dimensional versions of a~coarse-grained bead-spring model for the
  polymers, we investigate their structural properties and make a detailed
  comparison with structures of carbon nanotubes.
\end{abstract}

\maketitle

\section{Introduction}

In a recent computational study, it could be shown that flexible
polymers interacting with a wirelike substrate possess a barrellike
phase~\cite{vb1}. Optimally packed, the monomers form a cylindrical
polymer tube, reminiscent of a triangular lattice
which wraps around the wire. Depending on the competition between
steric constraints and monomer--substrate attraction, other structural
phases can also form.  These phases were also found in previous,
related studies of polymers interacting with
nano\-cylinders~\cite{binder1,srebnik1}.

Tubelike structures formed by atoms or molecules possess interesting
physical properties such as amazing mechanical stability, which make
them potential candidates for nanotechnological applications.
Recently, tin nanowires have been coated with atomic nanotube
structures in order to stabilize them for conducting superconductivity
experiments, i.e., protect them from shape fragmentation as well as
from oxidation~\cite{tombros08nanolett}. The understanding of the
wetting behavior of atomic nanotubes with polymeric materials has been
claimed to be the key to carbon nanotube--polymer
composites~\cite{tran08nanolett}. Biological cells require a stable
cytoskeleton which consists of tubelike myosin fibers.

The most prominent examples of tubes on atomistic scales are carbon
nanotubes~\cite{Iiji,roche07rmp} which can be thought of as ``rolled-up'' and
``zipped'' sheets of graphene, sharing its hexagonal honeycomb lattice
structure and sp$^2$ hybridized atoms~\cite{nano1,nano2,nano3,dres}.
Specifically, single-walled carbon nanotubes (SWCNTs) have been
extensively studied on different levels of approximation. While
nanotube models are typically based on continuum approximations, it
has recently been shown that their atomistic nature is
crucial for correct estimation of nanotube
\hbox{parameters}~\cite{Huang,pine}.

As well as carbon nanotubes, boron and boron nitride
tubes have also been created~\cite{ciu1} and modeled. A
review of boron tube modeling is given in Ref.~\cite{lee1}. The
salient differences between boron and carbon tubes are that the boron
tubes form a triangular lattice structure (as do the polymer tubes)
but the boron tubes appear to require either puckering, substitution
with nitrogen or regular vacant sites for stability, unlike
both polymer and carbon tubes. Thus the polymer tubes share one
feature~-- the underlying lattice~-- with single-walled boron
nanotubes (SWBNTs) and another~-- non-buckled, translationally
invariant surfaces~-- with the carbon tubes.

From a formal point of view the hexagonal nanotube atomic lattice is
dual~\cite{Essam} to the triangular lattice, suggesting there may be a
deeper connection.  This has been extensively explored for idealized
single-walled boron nanotubes~\cite{lee1}, but the buckling or regular
vacancies in real boron tube structures complicate a precise modeling.
The polymer tubes we investigate in this study are complete, unbuckled
triangulations of single-walled tubes and thus we can directly adopt
the theory introduced in Ref.~\cite{lee1} for idealized boron tubes to link
our results for polymer tubes with known atomic boron and carbon
nanotube structures.

This paper is structured as follows. The description of the hybrid
polymer--wire model leading to monolayer polymer tube conformations
for certain parameters, and a summary of our previous findings on
those systems, is given in Sect.~\ref{sec:model}. In
Section~\ref{sec:revCN} we present details of typical nanotube
configurations and quantify their characterization. Since the
  correct treatment of discrete tube structures is indispensable for
  our discussion and was introduced quite recently, we also review the
  polyhedral model for the description of ideal nanotubes in detail.
  In Sections~\ref{sec:theo2} and~\ref{sec:sim} we will present a
  detailed discussion of our mappings and simulations based on Monte
  Carlo simulations in the full three dimensional space
  (Sect.~\ref{sec:simA}). These simulations indicate that it is indeed
  adequate to restrict the investigation to polymers on cylindrical
  surfaces (Sect.~\ref{sec:simB}) in order to introduce a precise
  classification of polymer tubes. We show that while certain
crucial differences are present, there is a deep similarity between
atomistic nanotubes and polymer tubes. A~summary of observations and
conclusions will complete the paper.

\section{Structural properties of polymer tubes}
\label{sec:model}

\subsection{Polymer--wire model}

For our study of polymer tubes, we employ a coarse-grained hybrid model of a flexible,
elastic polymer interacting with an attractive stringlike nanowire. We found
recently that such a system possesses a conformational 
phase, in which tubelike monolayer structures spontaneously form~\cite{vb1}. In
our model,
pairs of monomers interact via a truncated and shifted Lennard-Jones
(LJ) potential
\begin{equation}
V_{\rm LJ}^{\rm mod}(r_{ij})=V_{\rm LJ}(\min(r_{ij},r_c))-V_{\rm LJ}(r_c)
\end{equation}
with the standard form of the LJ potential
\begin{equation}
V_{\rm LJ}(r_{ij})=4\epsilon[(\sigma/r_{ij})^{12}-(\sigma/r_{ij})^{6}], 
\end{equation}
where $r_{ij}$ denotes the distance between the $i$th and $j$th
monomer.  We set the respective intrinsic energy and length scales to
$\epsilon=1$ and $\sigma=2^{-1/6}r_0$ with the minimum-potential
distance $r_0=1$.  The cutoff is chosen to be $r_c=2.5\sigma$.
Covalently bonded adjacent monomers in the linear polymer chain
interact via the finitely extensible nonlinear elastic (FENE)
potential, which has the form~\cite{FENE,binder3}
\begin{equation}
V_{\rm FENE}(r_{i\,i+1})=-\frac{K}{2}R^2\ln\left\{1-[(r_{i\,i+1}-r_0)/R]^2\right\}.
\end{equation}
Its minimum coincides, by construction, with $r_0$ and diverges for
$r\rightarrow r_0\pm R$. We set $R=0.3$ and $K=40$.

The interaction of the polymer with the wire is modeled by the
potential
\begin{equation}
V_\mathrm{string}(r_{\perp;i})=\pi\, a\epsilon_f\left(\frac{63}{64}
\frac{\sigma_f^{12}}{r_{\perp;i}^{11}}-\frac{3}{2}\frac{\sigma_f^{6}}{r_{\perp;i}^5}\right),
\label{eq:string}
\end{equation}
where $\sigma_f$ and $\epsilon_f$ are the monomer--wire interaction
parameters and $r_{\perp;i}$ is the distance of the $i$th monomer
perpendicular to the wire. We scale the potential such that its
minimum value is $-1$ at $r_{\perp}^\mathrm{min}$ for $\epsilon_f=1$
and $\sigma_f=1$, in which case $a\approx 0.528$; see
also~\cite{vogelCPC}. The effective thickness of the string,
$\sigma_f$, is related to the minimum distance
$r_{\perp}^\mathrm{min}$ of the monomer--wire potential via
\begin{equation}
r_{\perp}^\mathrm{min}(\sigma_f)=\left({693}/{480}\right)^{1/6}\sigma_f\approx1.06\,\sigma_f\,.
\label{eq:sigma}
\end{equation}
Eventually, the total energy of the polymer interacting with the wire is given by
\begin{equation}
E=\kern-2mm\sum_{i,\,j=i+1}^N\kern-2mm V_{\rm LJ}^{\rm mod}(r_{ij})+\sum_{i=1}^{N-1} V_{\rm FENE}(r_{i\,i+1})+\sum_i^N V_\mathrm{string}(r_{\perp;i}).
\end{equation}
In order to identify structural properties of low-en\-ergy adsorbed polymer conformations,
we employed stochastic minimization techniques based on generalized-ensemble Monte Carlo
sampling strategies such as the energy-landscape paving method~\cite{elp}, 
multicanonical sampling~\cite{muca}, 
and the Wang--Landau method~\cite{wl}.

\subsection{Structural phases of the polymer--wire system}

\begin{figure}[b!]
\begin{center}
\includegraphics[width=50mm,clip]{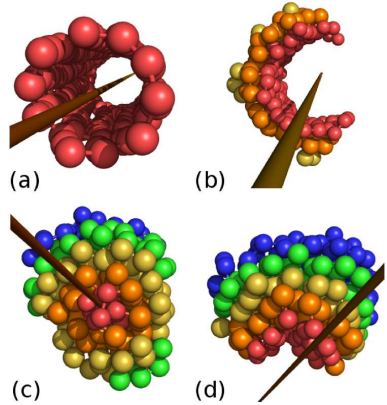}
\end{center}
\caption{\label{fig:1} Exemplified images of representative low-energy
conformations in the four structural
phases (see text) of a~polymer
with 200 monomers interacting with an attractive nanowire. The parameter settings 
are: (a)~$\epsilon_f=5.0$, $\sigma_f=1.5$ (in structural phase B), 
(b) $\epsilon_f=3.5$, $\sigma_f=2.5$ (C), 
(c)~$\epsilon_f=2.0$, $\sigma_f=0.5$ (Gi), 
(d) $\epsilon_f=1.0$, $\sigma_f=1.5$ (Ge).}
\end{figure}

In close analogy to a recent study of a bead-stick polymer interacting with a
nanowire~\cite{vb1}, the full spectrum of structural phases can be revealed.
Depending on the wire attraction strength $\epsilon_f$ and its
effective thickness $\sigma_f$, spherically globular or rather
extended conformations dominate. Although in this paper we are interested in the tubelike
structural phase of the polymer--wire system only, let us briefly review all
phases identified. In Fig.~\ref{fig:1}, representative
examples of low-energy conformations in the different phases are depicted.

If the value of $\epsilon_f$ is small enough that monomer--monomer
contacts are energetically more favorable than contacts with the
substrate, the lowest-energy conformations are compact spherical
globules.
Since the wire is always attractive, the number of
monomer--substrate contacts is also maximized, such that the globular
structures inclose the wire [phase ``Gi'' (globular inclosed), see
Fig.~\ref{fig:1}(c)].  If, on the other hand, the length scale of the
monomer--wire interaction exceeds the length scale of the pairwise
LJ-interaction among two monomers, the wire is pushed outward and the
globule is simply attached to the the wire [this phase is called
``Ge'' (globular excluded), see Fig.~\ref{fig:1}(d)]. Starting from
phase Ge and increasing the energy scale of the attraction to the
wire, the morphology of conformations changes. The wire-attached
spherically symmetric globules ``melt'' along the wire axis and reach
what is called the ``clamshell phase'' C, see Fig.~\ref{fig:1}(b). The
spherical morphology is broken and the polymer starts wrapping around
the wire in order to increase the energetically favored contacts with
the substrate.

If the string thickness is reduced below the corresponding threshold
value, the clamshells turn to ``barrels'' and the structural phase B
is reached. The same scenario occurs, when approaching from the Gi phase
and passing the transition point, where the energy scale of the
monomer--substrate attraction is sufficiently large compared to the
intrinsic polymer energy scale of nonbonded LJ-interactions to allow
for an increase of monomer--substrate contacts at the expense of
monomer--monomer contacts. Since additional contacts with the wire can
only be formed along the wire axis, the polymer forms compact tubelike
structures in this phase, see Fig.~\ref{fig:1}(a). 

For our discussion of similarities of the polymer--wire system with
carbon nanotubes, we will refer below only to the monolayer polymer
tube structures formed in phase B.  Before embarking on the comparison
of polymer tubes and carbon nanotubes, we briefly review
relevant geometrical properties of SWCNTs.

\section{Single-walled nanotubes}
\label{sec:revCN}

Nanotubes are typically considered as rolled-up planar atomic sheets.
A single-walled carbon nanotube (SWCNT) is commonly pictured as a
zipped monolayer graphene sheet (although, more detailed approaches
exist~\cite{bera11arxiv}), crystallized on a hexagonal (honeycomb)
lattice.  All carbon atoms have the same distance from the tube axis
and thus reside on a cylinder surface. In the original, conventional
model, the SWCNT lattice was supposed to entirely cover a cylinder,
thereby assuming curved hexagonal plates and bonds, neglecting the
discrete nature of the lattice~\cite{dres}. In a more realistic
polyhedral decomposition approach, this is corrected by formulating
constraints regarding C-C bond lengths and angles between
them~\cite{lee1,lee2}.

Single-walled boron nanotubes (SWBNTs) possess an underlying
triangular lattice structure and have also been described by
cylindrical mappings~\cite{lee1,lee3}. Energetically favored tube
structures possess holes or are puckered, in which case the atoms do
not lie on a surface of a single cylinder~\cite{kunst1,tian1}. Other,
non-regular structures, have also been considered in theoretical
studies of SWBNTs~\cite{wang09cpc}.

The triangular single-walled polymer tube (SWPT) structures we find in
the barrel phase of a polymer adsorbed at a nanowire exhibit strong
similarities with ideal cylindrical SWBNTs. Thus, for its geometrical
description the ideal polyhedral model for \hbox{SWBNTs} with equal bond
lengths~\cite{lee3} can easily be adopted. Since the polymers tend to
form highly regular tube structures, they are interesting candidates
for carbon nanocomposites of polymers and SWCNTs. For these reasons,
it is instructive to discuss the relationship of SWCNTs and SWPTs in
the following.

\subsection{The conventional view on carbon nanotubes}

\begin{figure}[t]
\includegraphics[width=\textwidth,clip]{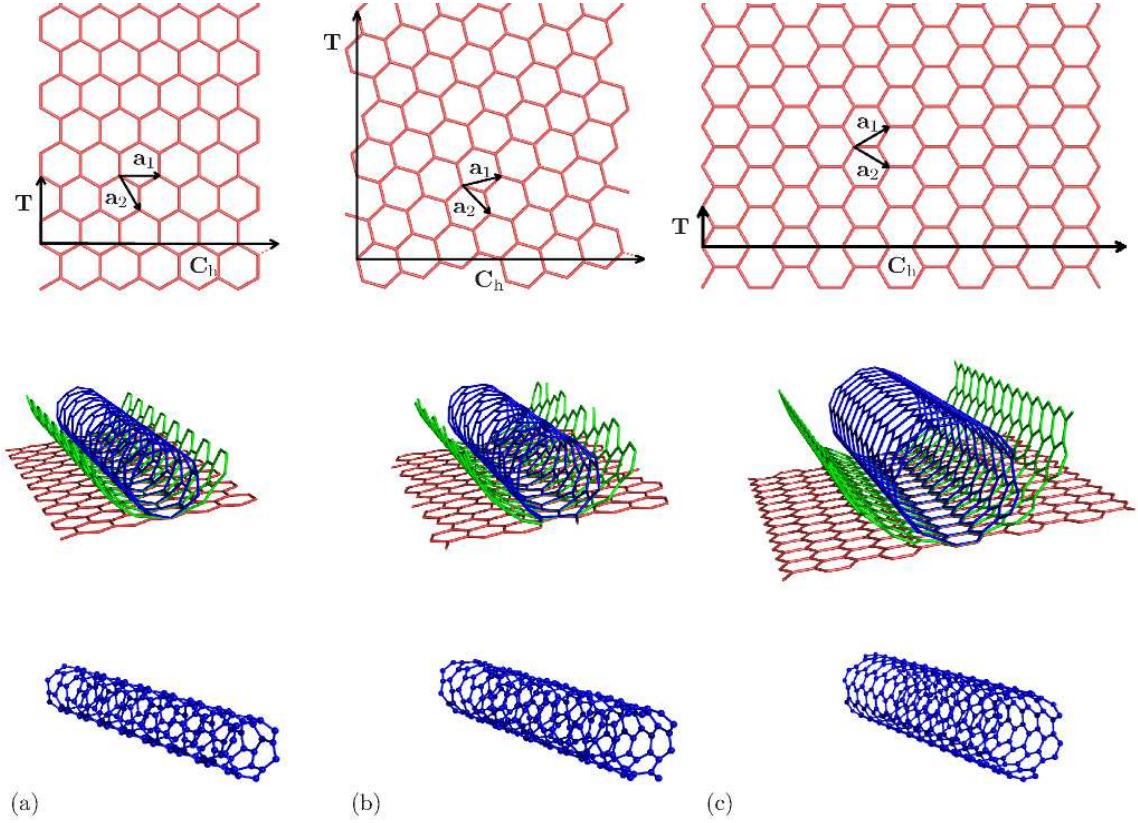}
\caption{Examples of carbon nanotube structures: (a) zigzag (6,0), (b)
  chiral (6,2), and (c) armchair (6,6) conformation. In the top row,
  unzipped and unrolled planar representations are shown, the middle
  row illustrates the zipping and at the bottom, the actual tube
  structures are visualized.}
\label{fig:2-3tubes}
\end{figure}

In the unzipped, conventional representation, the chiral or wrapping
vector $\mathbf{C}_\mathrm{h}$ pointing from any lattice site to its
next copy (see Fig.~\ref{fig:2-3tubes}) uniquely characterizes any SWCNT
structure. The wrapping vector and the translational vector
$\mathbf{T}$, perpendicular to $\mathbf{C}_\mathrm{h}$, span the unit
cell. It is convenient to introduce lattice vectors $\mathbf{a}_1$ and
$\mathbf{a}_2$ (see Fig.~\ref{fig:2-3tubes}), such that
$\mathbf{C}_\mathrm{h}=n\,\mathbf{a}_1+m\,\mathbf{a}_2$ and
$\mathbf{T}=[(n+2m)/d]\,\mathbf{a}_1-[(2n+m)/d]\,\mathbf{a}_2$, where $d$
is the greatest common divisor of $n+2m$ and $2n+m$. Hence, the two
integers $n$ and $m\leq n$, usually written in the vector form
($n,m$), are sufficient to differentiate between SWCNT structures.

The wrapping orientation is defined by the characteristic wrapping or
chiral angle $\theta$ between $\mathbf{a}_1$ and
$\mathbf{C}_\mathrm{h}$, i.e.,
\begin{equation}
\cos\theta_{\mathrm{conv}}^{(n,m)}=(2n+m)/2\sqrt{n^2+m^2+nm}\,.
\label{eq:theta_conv}
\end{equation}
Hence, $\theta$ can take values between $0$ and $30^\circ$. The corresponding
limiting tube conformations for a given value of $n$ are usually
called zigzag (for $m=0$) or armchair conformation (for $m=n$). In Fig.~\ref{fig:2-3tubes}, different
visualizations of ($6,m$) zigzag, chiral, and armchair carbon nanotube
structures are shown.

In this conventional approach~\cite{dres}, SWCNTs were assumed to be planar sheets of the
$(n,m)$ unit cell wrapped around a cylinder and continued along the
central axis. Then, the length of $\mathbf{C}_\mathrm{h}$ corresponds
to the circumference $L_\mathrm{conv}^{(n,m)}$ of this cylinder. Since
\begin{equation}
|\mathbf{C}_\mathrm{h}|=L_\mathrm{conv}^{(n,m)}=\sqrt{3}\,l_\mathrm{CC}\,\sqrt{n^2+m^2+nm}\,,
\label{eq:Lconv}
\end{equation}
where $l_\mathrm{CC}\approx 1.42\text{\AA}$ corresponds to the C--C bond length, the radius of the
zipped SWCNT is given by
\begin{equation}
r_\mathrm{conv}^{(n,m)}=L_\mathrm{conv}^{(n,m)}/2\pi\,.
\label{eq:convrad}
\end{equation}
In this simple cylindrical mapping, the bond length is not
conserved and discrete curvature effects are not correctly taken into
account.  However, the flatter the surface, i.e., the larger $n$, the
more accurate this estimate is.

For structural investigations of realistic SWCNTs, this difference is
not of particular relevance. However, for our subsequent comparison
and discussion of the relationship between triangular SWPTs
and hexagonal SWCNT shapes, the (typically small) deviations must be
taken into consideration, because the structural characterization on
the basis of the $(n,m)$ vector depends sensitively on this.

Assuming that the SWCNT forms under the constraints of conserved C--C
bond\break{} lengths, which are not necessarily all equal~\cite{budyka1}, a
discrete SWCNT model can be derived. The obtained geometric tube
structures are in good correspondence with ab initio predictions and
molecular dynamics relaxation~\cite{lee1,lee4}. This model is also
applicable to ultra-small nanotubes which resemble
nanowires~\cite{lee5}.

\subsection{Polymer tubes and the polyhedral model for nanotubes}

In the tube phase, polymers attracted by a thin wire form compact
conformations. Because of the elasticity and flexibility of the
polymer model considered here, monomers are optimally packed in a
triangular arrangement. Unzipping such a polymer tube yields a regular
triangular lattice, whose lattice vectors are identical with
$\mathbf{a}_1$ and $\mathbf{a}_2$ introduced earlier for the
definition of the wrapping vector $\mathbf{C}_h$ of the SWCNTs.
Consequently, the $(n,m)$ notation for the characterization of SWCNTs
can also be implemented to characterize the SWPTs.

The polymer tubes we find in the monolayer barrel phase of our
polymer--nanowire adsorption model can be well described by the ideal
boron nanotube model~\cite{lee1,lee3}, where all bonds are considered
to have an identical length $l_\mathrm{BB}$. This assumption
corresponds well to the definitions of the length scales of bonded and
nonbonded interactions in the FENE polymer model used in our study.
Thus, the systems can easily be mapped onto each other by the
replacement $l_\mathrm{BB} \leftrightarrow r_0$. The polymers form
monolayer tubes, if the adsorption strength overcompensates optimal
three-dimensional nearest-neighbor packing of the monomers. If the
adsorption strength is reduced to the extent that intrinsic attraction
becomes competitive, the topologically two-dimensional monolayer is
given up in favor of a double-layer structure extending into the third
dimension. The polymer undergoes a topological transition, but keeps a
barrellike form~\cite{vb1}. In the following, we only consider
monolayer polymer tubes, as only in this conformational phase the
analogy to carbon and (idealized) boron nanotubes is apparent.\vadjust{\break}

\begin{figure}
\begin{center}
\includegraphics[width=55mm,clip]{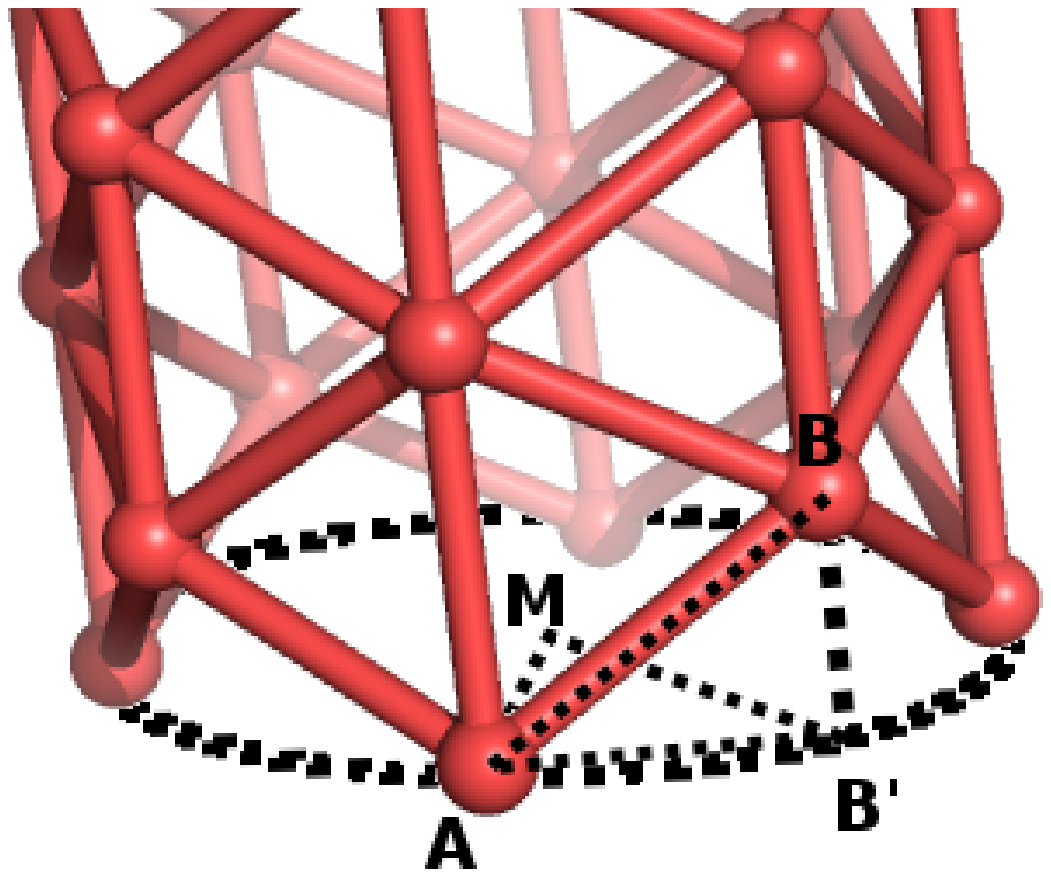}
\end{center}\vspace*{-4mm}
\caption{Definitions of the angles $\psi$ and $\theta$, shown
  exemplarily using a part of a (4,4) SWPT.
  $2\,\psi=\angle(\mathbf{AMB^\prime})$;
  $\theta=\angle(\mathbf{BAB^\prime})$.
\label{fig:3-angle_def}}
\end{figure}

Adopting the equations from the idealized boron nanotube
model~\cite{lee3} and the chiral notation $(n,m)$ from atomic
nanotubes, the radius of an $(n,m)$ polymer tube is given by
\begin{equation}
\label{eq:polyrad}
r_{\mathrm{poly},\vartriangle}^{(n,m)}=\frac{r_0}{2}\frac{\cos \theta^{(n,m)}}{\sin\psi^{(n,m)}}\,,
\end{equation}
where $r_0$ is the equilibrium monomer--monomer distance~\cite{footnote1} and the corrected wrapping angle
$\theta^{(n,m)}$ is given by
\begin{equation}
\cos^2\theta^{(n,m)}=
\frac{n(n+2m)\sin^2\psi^{(n,m)}}{(n+m)^2\sin^2\psi^{(n,m)}-m^2\sin^2(\psi^{(n,m)}+\xi^{(n,m)})}\,.
\label{eq:corrTheta}
\end{equation}
The angles $\psi^{(n,m)}$ and $\xi^{(n,m)}=(n\psi^{(n,m)}-\pi)/m$ are
defined by projections of atom positions upon a circular slice perpendicular to
the tube axis~\cite{lee3}. They are obtained by solving the transcendental equation
\begin{equation}
\label{eq:psi}
0=(n^2-m^2)\,\sin^2\left(\psi^{(n,m)}+\xi^{(n,m)}\right)
-n(n+2m)\,\sin^2\xi^{(n,m)}
+m(2n+m)\,\sin^2\psi^{(n,m)}\,.
\end{equation}
See also Fig.~\ref{fig:3-angle_def} for definition of the angles $\psi$ and $\theta$.

Analogously one can formulate equations for $\theta^{(n,m)}$ and
$r_{\mathrm{poly},\hexagon}^{(n,m)}$ for the polyhedral model for carbon
nanotubes with fixed bond lengths, crystallized in a curved honeycomb
lattice structure. In fact, the wrapping angle $\theta^{(n,m)}$ for
SWCNTs is identical to that of SWPTs and \hbox{SWBNTs}, respectively, as the
base vectors $\mathbf{a}_{1,2}$ in the SWCNT structure correspond to
the bond vectors in triangular tubes.

The calculation of the polyhedral radius is more challenging, as the
honeycomb cell is not immune to shearing or tilting, as is a
triangular lattice. Indeed, the curvature induced by the wrapping
deforms the simple hexagonal cells, in contrast to triangular cells.
The first-order correction of the polyhedral radius for SWCNTs
is~\cite{lee4}:
\begin{equation}
r_{\mathrm{poly},\hexagon}^{(n,m)}=r_\mathrm{conv}^{(n,m)}
+\frac{\sqrt{3}\,\pi\,l_\mathrm{CC}\left[4(n^2+nm+m^2)^3-9n^2m^2(n+m)^2\right]}{64(n^2+nm+m^2)^{7/2}}
+\mathcal{O}(1/n^3)\,.
\label{eq:polyradCC}
\end{equation}

\noindent
For a concrete description of the limiting behavior of this model, we
start with the observation that
$\theta^{(n,m)}=\theta_\mathrm{conv}^{(n,m)}$
(cf.~Eqs.~(\ref{eq:theta_conv}) and~(\ref{eq:corrTheta})) for the two
special cases $n=m$ and $m=0$, i.e., for $(n,n)$- and $(n,0)$-tubes,
the conventionally calculated wrapping angles of $\theta=30^\circ$ and
$\theta=0^\circ$ are valid also in the polyhedral model. This leads to
an intuitive interpretation for the radius of triangular
$(n,0)$-tubes. From the vanishing wrapping angle follows that there
are bonds forming regular polygons with $n$ edges in planes
perpendicular to the tube orientation. Hence, the corresponding tube
radius is the circumradius of such a~polygon:
\begin{equation}
\label{eq:circumradn0}
r_{\mathrm{poly},\vartriangle}^{(n,0)}=\frac{r_0}{2\,\sin(\pi/n)}\,.
\end{equation}
Analogously, the radius of triangular $(n,n)$-tubes can be
directly calculated.

In this case, continuations of $\mathbf{a}_1$ or $\mathbf{a}_2$,
projected onto the wrapping vector (or, in the tube, the slice plane
perpendicular to the tube axis), will form a regular polygon with $2n$
edges. Since the angle between the original vectors $\mathbf{a}_{1,2}$
and their projections, i.e., the wrapping vector $\theta$, is
$\pi/6$, the lengths of the projected vectors are shortened by the
factor $\cos(\pi/6)=\sqrt{3}/2$. For the on-tube calculation of the
bond length between two monomers, we make use of the fact that the
projection angle in a triangular lattice does not change when zipping
the planar sheet to a tube:
\begin{equation}
\label{eq:circumradnn}
r_{\mathrm{poly},\vartriangle}^{(n,n)}=\frac{\sqrt{3}\,r_0}{4\,\sin (\pi/2n)}\,.
\end{equation}
Both, Eqs.~(\ref{eq:circumradn0}) and~(\ref{eq:circumradnn}), are of
course covered by Eq.~(\ref{eq:polyrad}), as Eq.~(\ref{eq:psi})
implies $\psi=\pi/2n$ and $\psi=\pi/n$ for $(n,0)$ and $(n,n)$ tubes,
respectively.

Note that it is not possible to apply similar arguments to calculate
the correct radii of $(n,0)$ and $(n,n)$ SWCNTs, because the hexagon
is not rigid, as the triangle is. Thus distances between the atoms in
the hexagon change when bending.  Indeed, for hexagonal honeycomb
tubes (SWCNTs) one finds, when projecting bonds to planes
perpendicular to the tube orientation, polygons with $3n$ and $2n$
edges for armchair and zigzag-tubes, respectively. But these polygons
are not regular anymore, in the armchair case, and the lengths of the
projections cannot be calculated in such a straightforward way, due to
minimal deformations caused by the curvature. However, applying such
approximations and corresponding generalizations to $(n,m)$ tubes lead
to the same qualitative results, such as the ($n,m$) sequence for
increasing radii, as shown below, for example. Quantitative deviations
from results of the polyhedral model are in the low per-mille range
(not shown).

Generally, although the numerical differences of
$r_\mathrm{conv}^{(n,m)}$ (Eq.~(\ref{eq:convrad}), with the respective
scale) and $r_\mathrm{poly}^{(n,m)}$ (Eqs.~(\ref{eq:polyrad})
and~(\ref{eq:polyradCC})) do not seem to be particularly striking, and
may not be distinguished in practice, in particular at finite
temperatures due to fluctuations of the bond lengths and the tube
itself, the physical consequences are actually important to notice.
Consider, for example, $(7,0)$ and $(5,3)$ SWCNTs. Using
Eq.~(\ref{eq:convrad}), both would share the same radius
$r_\mathrm{conv}^{(7,0)}=r_\mathrm{conv}^{(5,3)}=\sqrt{3}\,l_\mathrm{CC}\,\sqrt{7}/2\pi$
and could not be distinguished using this quantity. However, the
structures of the corresponding tubes are completely different, and in
consequence, physical, i.e., nanoelectronic, nanooptic and other
material properties~\cite{bernholc,kim,dres,dres2}, are in general
different as well. Hence, $r_\mathrm{conv}^{(n,m)}$ is not suitable to
parametrize SWCNTs, whereas $r_\mathrm{poly}^{(n,m)}$ can uniquely be
associated to any SWCNT structure. A more detailed numerical
analysis of the deviations between the conventional and the polyhedral
approach is given in Ref.~\cite{vb_athens11}.

However, for the following discussion of similarities of SWCNTs and
SWPTs, significant precision is required and therefore it is
necessary to take into account these \hbox{differences}.

\section{Mapping between carbon nanotubes and polymer tubes}
\label{sec:theo2}

Since the triangular lattice formed by an unzipped SWPT is obtained
from a Voronoi construction of the hexagonal lattice formed by the
carbon atoms in an unzipped SWCNT, it is appealing to investigate the
mapping between these different systems. This is not only
mathematically interesting, but might also
have technological consequences for the design of particularly stable
polymer coated carbon nanotubes or other nanohybrid structures
including SWCNTs and complex
molecules~\cite{gao03ea,panhuis03jpcb,ehli06jacs,tran08nanolett,ehli09nat,srebnik1,caddeo10jpcc,tallury10jpcb}.

However, as we have discussed in the previous section, it is necessary
to examine geometrical considerations on the tube itself and not on its
unzipped form. This is not trivial due to the curvature and discrete
nature of those systems, which affects, for example, the bending
angles between the respective carbon atoms or monomers such that they
generally differ in the unzipped planar shape and in the tube
structure.

\begin{figure}[t]
\includegraphics[width=\textwidth,clip]{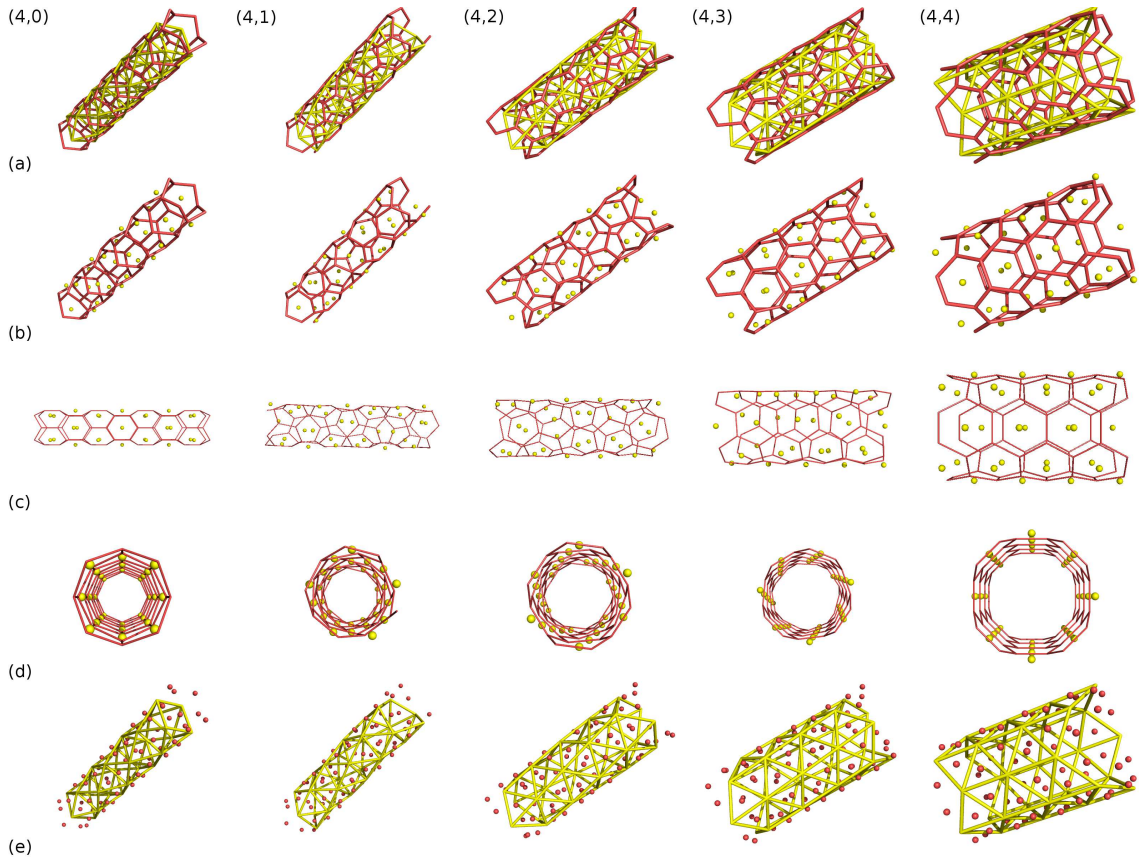}
\caption{Different visualizations of $(4,m)$ nanotubes
  with (l. to r.)  $ 0 \le m \le 4 $.  Hexagonal SWCNT (red online)
  and the corresponding ideal triangular tubes (yellow online) are
  shown from different viewpoints and combinations of lattice points
  (atomic positions) and edges (bonds). The data is the same for each
  column ($m$ value) with all lattice points always shown, but edges
  only in some cases specifically (a) both, (b), (c) and (d) bonds of
  SWCNTs only and (e) bonds of triangular tubes only.  }
\label{fig:4}
\end{figure}

Exemplified for $(4,m)$ tubes, Fig.~\ref{fig:4} shows in various
types of visualization the general construction principle of
triangular polymer tubes out of honeycomb SWCNT structures. A monomer
of the polymer chain is placed in the center of each ``hexagonal''
plaquette on the tube, i.e., at the position of the vertices of the
associated Voronoi graph of the hexagonal lattice points in flat
(unzipped) space, and then the monomer is shifted along the axis
perpendicular to the tube axis to the correct distance
$r_{\mathrm{poly},\hexagon}^{(n,m)}$ from the tube axis.
Fig.~\ref{fig:4}\,(b) and~(c) enable one to sense the generally
non-trivial geometrical structure of SWCNTs.  In
Fig.~\ref{fig:4}\,(d) the helical aspect of the corresponding
triangular tubes, to which we will return later, is \hbox{emphasized}.

\begin{figure}
\begin{center}
\includegraphics[width=92mm,clip]{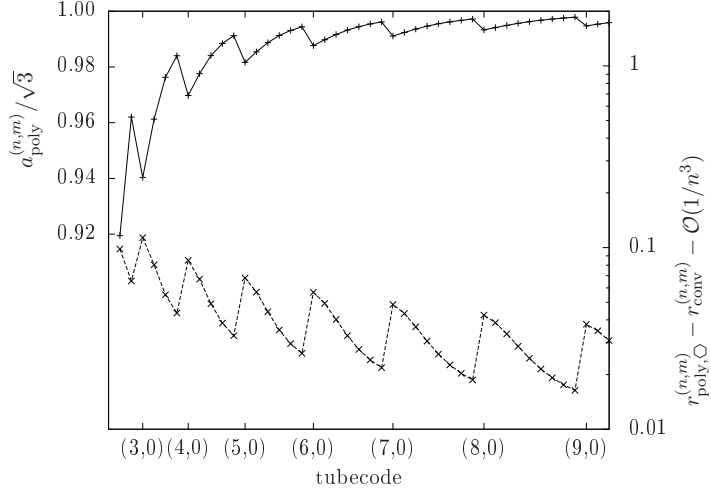}
\end{center}
\caption{Solid graph, left scale: Normalized scaling of lengths in
  SWCNTs compared to SWPTs in the polyhedral model. The upper bound of
  the plot corresponds to the limiting value of
  $a^{(n,m)}_\mathrm{poly}/\sqrt{3}$ for $n\to\infty$. Dashed graph,
  right (logarithmic) scale: numerical value of the first order
  correction term to $r_\mathrm{conv}^{(n,m)}$ in
  Eq.~(\ref{eq:polyradCC}) for SWCNTs in the polyhedral model. The $x$
  axis is ordered with respect to increasing $n$ and $m$, i.e.,
  between two tics marked $(n,0)$ and $(n+1,0)$ tubes, all $(n,m)$
  tubes with increasing $0<m\leq n$ are located.}
\label{fig:5}
\end{figure}

The length scale in the triangular tube is obviously different from
that in the SWCNT. The resulting scaling of bond lengths is of
particular interest. Equivalently, one can ask for the radius of an
$(n,m)$ polymer tube, with a monomer--monomer bond length scale $r_0$
equal to $l_\mathrm{CC}$. As known, the scaling factor in the
conventional planar representation is $a_\mathrm{conv}=\sqrt{3}$
independently of $n$ or $m$, which is hence the limiting case for
$n\to\infty$ in the tube geometry. For small $n$ and $m$,
\begin{equation*}
\frac{a^{(n,m)}_\mathrm{poly}}{\sqrt{3}}:=\frac{r^{(n,m)}_{\mathrm{poly},\hexagon}(l_\mathrm{CC})}{\sqrt{3}\,r^{(n,m)}_{\mathrm{poly},\vartriangle}(r_0=l_\mathrm{CC})}\quad\stackrel{n\to\infty}{\longrightarrow}\;1
\end{equation*}
is plotted in Fig.~\ref{fig:5} (solid line, ``$+$'' symbols, left scale).

The influence of the first correction term to
$r_\mathrm{conv}^{(n,m)}$ in Eq.~(\ref{eq:polyradCC}) is connected to
this \hbox{scaling}. The term is also plotted in Fig.~\ref{fig:5} (dashed
line, ``$\times$'' symbols, right scale) and can be considered to be
an estimate of the error made by applying Eq.~(\ref{eq:convrad}) for
the calculation of the SWCNT radius. This error can be bigger than the
differences between radii of two SWCNTs of different type~\cite{vb_athens11}.

\section{Computer simulations of polymer tube structures}
\label{sec:sim}

In the following, we present and discuss results from Monte Carlo
minimizations of the bead--spring polymer--wire system introduced in
Sect.~\ref{sec:model} for strong wire attraction, such that the ground
states form SWPTs. Compared to the model used in
Ref.~\cite{vb1}, we introduced extendible bonds and adjusted the
equilibrium distance of the Lennard-Jones interactions such, that it
coincides with the equilibrium length of the FENE bond potential. This
simplifies the simulations and the bond-length flexibility limits, at
this point, the occurrence of defects.

\subsection{Simulation in the full conformational space}
\label{sec:simA}

We perform simulations using generalized-ensemble Monte Carlo
techniques~\cite{elp,muca,wl} to search for low-energy configurations.
We propose new structures through local updates of the Cartesian
monomer coordinates, global spherical-cap updates~\cite{mb05pre},
slithering snake moves, and bond-exchange moves~\cite{schnabel10cpc}.

\begin{table}
{\footnotesize{
\begin{tabular*}{\textwidth}{@{\extracolsep{\fill}}l|cccc|ccc}
input & \multicolumn{4}{c}{measured from lowest-energy state} & \multicolumn{3}{c}{calculated using polyhedral model} \\
$\sigma_{{f}}$ & $r$ & $2\psi$ in ${}^\circ$ & $\theta$ in ${}^\circ$ & $(n,m)$ & $r_{\mathrm{poly},\vartriangle}^{(n,m)}$ & $2\psi_{\mathrm{poly},\vartriangle}^{(n,m)}$ in ${}^\circ$ & $\theta_{\mathrm{poly},\vartriangle}^{(n,m)}$ in ${}^\circ$\\
\hline
0.47 & $0.504\pm0.002$ & $131.9\pm2.6$ & $19.1\pm0.7$ & (2,1) & 0.51962 & 131.8 & 18.43 \\
0.53 & $0.566\pm0.002$ & $120.0\pm0.8$ & $0.0\pm0.4$  & (3,0) & 0.57735 & 120.0 & 0.00  \\
0.56 & $0.597\pm0.002$ & $89.9\pm0.8$  & $30.7\pm0.5$ & (2,2) & 0.61237 & 90.0  & 30.0  \\
0.61 & $0.646\pm0.002$ & $98.0\pm0.9$  & $13.3\pm0.6$ & (3,1) & 0.64526 & 97.7  & 13.6  \\
0.65 & $0.692\pm0.002$ & $90.0\pm0.4$  & $0.0\pm0.3$  & (4,0) & 0.70711 & 90.0  & 0.00  \\
0.69 & $0.732\pm0.003$ & $76.5\pm0.6$  & $23.5\pm0.3$ & (3,2) & 0.74313 & 76.3  & 23.3  \\
0.74 & $0.783\pm0.003$ & $77.4\pm0.5$  & $10.6\pm0.4$ & (4,1) & 0.78561 & 77.4  & 10.7  \\
0.78 & $0.831\pm0.003$ & $72.1\pm0.6$  & $0.0\pm0.4$  & (5,0) & 0.85065 & 72.0  & 0.00  \\
0.81 & $0.858\pm0.003$ & $60.0\pm0.3$  & $30.1\pm0.5$ & (3,3) & 0.86603 & 60.0  & 30.0  \\
0.83 & $0.880\pm0.004$ & $64.7\pm0.4$  & $19.0\pm0.4$ & (4,2) & 0.88462 & 64.6  & 19.0  \\
0.88 & $0.932\pm0.004$ & $64.0\pm0.3$  & $8.8\pm0.4$  & (5,1) & 0.93259 & 64.0  & 8.84
\end{tabular*}
}}
\caption{Characteristics of polymer tubes formed by FENE-polymers adsorbed to a string in the polymer--wire model. The measured quantities are average values from local measurements at single monomers in the lowest-energy structures, see the text for details.}
\label{tab:1}
\end{table}

\begin{figure}[b!]
\begin{center}
\includegraphics[width=.85\textwidth,clip]{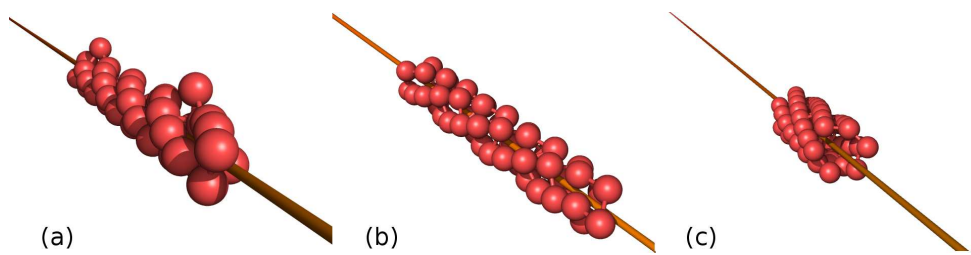}
\end{center}
\caption{Low-energy tube structures of the simulated
    polymer--wire model. The structures were found for different
    $\sigma_f$ (cp. Table~\ref{tab:1}): (a) $\sigma_f=0.47$. Slightly
    excited state, two monomers do not belong to the monolayer around the
    wire. The monomers in that closest layer form a (2,1) tube. (b)
    $\sigma_f=0.61$. Putative ground-state forming a (3,1) tube around
    the wire. (c) $\sigma_f=0.83$. Putative ground-state forming a
    (4,2) tube. All shown polymers consist of $N=48$ monomers.}
\label{fig:6-struct3D}
\end{figure}

The results of simulations with polymers of the length $N=32$ are
summarized in Table~\ref{tab:1}. In all our simulations, we set
$\epsilon_f=5.0$, i.e., we simulate structures in the monolayer barrel
region~'B'{\,}~\cite{vb1}. The input to the simulation is $\sigma_f$,
i.e., the effective thickness of the wire, is given in the first
column. In the following columns, average values
of measured local observables are shown. The tube radius $r$ is, for
example, measured as $r=\sum_{i=1}^Nr_i/N$, with $r_i$ being the
perpendicular distance of the $i$th monomer from the center of the
string. The small variance indicates that the monomers are located on
a cylinder surface, in fact. Definitions of the characteristic angles
$\psi$ and $\theta$, which are also measured locally at each monomer
and then averaged, are given in Fig.~\ref{fig:3-angle_def}. We convinced
ourself by simulating all structures at different chain lengths that
the results are independent of the actual choice of $N$. By applying
the polyhedral model for triangular tubes, an $(n,m)$ tube code, given
in the fifth column, can uniquely be assigned to each pair
$(\psi,\theta)$.  Finally, in the last three columns, the
corresponding calculated geometric observables using the polyhedral
model are given, which agree perfectly with the measured ones. The
radius does not match exactly though, which is due to the fact that we
did not use the calculated values from the model as input for the
simulations, but $\sigma_f$ in steps of $0.01$ to avoid any potential
bias.  However, the flexibility of the bonds allows the compensation
of these small deviations without causing defects in the ground-state
structures.  We will comment on that in more detail in the next
section. After all, we find, that the polyhedral model for boron
nanotubes~\cite{lee1} is suitable to describe the completely adsorbed
polymer ground-state structures of a simple polymer--wire model.  Some
example structures are visualized in Fig.~\ref{fig:6-struct3D}.

\subsection{Simulations on a cylinder surface}
\label{sec:simB}

Since we find in the perfect monolayer tube conformations all monomers
at the same distance from the tube axis, i.e., fluctuations of the
local radius of the monolayer tubes were in fact vanishing, it is
useful to directly simulate the polymer on a cylinder surface with
fixed radius~$r$. Following this idea, we randomly initialize
conformations in such a way that each monomer has the same distance
$r$ from the string. We propose new structures by local updates and
slithering snake moves in cylindrical coordinates keeping $r$
untouched. This reduces the conformational space significantly and
will allow us to refine our results shown above and determine, for
example, intervals of radii with \hbox{stable} ground-state conformations.
Furthermore, longer chains and hence larger tube radii could be
studied. Finally, as an outlook to subsequent work, defects can be
investigated in detail. See Sect.~\ref{sec:OA} below, for an example.

\begin{table}[b!]
{\footnotesize{
\begin{tabular*}{\textwidth}{@{\extracolsep{\fill}}lcr|c|ccc|cc}
\multicolumn{3}{c}{Calculated polyhedral} &
\multicolumn{4}{c}{Simulation on Cylinder surface (``2D'')} &
\multicolumn{2}{c}{Corresponding} \\
\multicolumn{3}{c}{polymer tube}& input & \multicolumn{3}{c}{output (Lowest-energy state)} &
\multicolumn{2}{c}{SWCNT} \\
($n,m$) & $r_{\mathrm{poly},\vartriangle}^{(n,m)}$ & $2\psi$ in ${}^\circ$ & $r$ & type & $2\psi$ in ${}^\circ$ & $\theta$ in ${}^\circ$ & $\theta$ in ${}^\circ$ & type \\
\hline
(1,1) &	       &        & 0.425            & (1,1)       & $181\pm6$ & $30.4\pm0.5$                    & 30.00& (1,1) \\[-.5pt]
      &        &        & 0.436\ldots0.447 & Double-helix$^\ast$\hspace*{-2.7mm} &  & ($29.2$\ldots$27.6$)$\pm0.7$  &        &       \\[-.5pt]
(2,0) &        & 	& 0.457\ldots0.466 & (2,0)        & $180\pm2$ &$0.0\pm1.2$ &  0.00& (2,0) \\[-.5pt]
(2,1) &	0.51962& 131.810& 0.477\ldots0.532 & Triple-helix$^\ast$\hspace*{-2mm} & $132\pm2$ & ($20.8$\ldots$17.5$)$\pm1.0$  & 18.43& (2,1) \\[-.5pt]
(3,0) &	0.57735& 120.000& 0.553\ldots0.574 & (3,0)$^\ast$       & $120\pm0.8$ & $0.0\pm0.5$                     &  0.00& (3,0) \\[-.5pt]
(2,2) &	0.61237& 90.0000& 0.585\ldots0.617 & (2,2)$^\ast$       & $90\pm1$ & ($31.3$\ldots$29.6$)$\pm0.3$  & 30.00& (2,2) \\[-.5pt]
(3,1) &	0.64526& 97.7431& 0.627\ldots0.670 & 4-helix     & $98\pm1$  & ($13.9$\ldots$12.6$)$\pm0.7$  & 13.57& (3,1) \\[-.5pt]
(4,0) &	0.70711& 90.0000& 0.680\ldots0.712 & (4,0)       & $90.0\pm0.7$ & $0.0\pm0.5$                     &  0.00& (4,0) \\[-.5pt]
(3,2) &	0.74313&  76.3120& 0.723\ldots0.755 & 5-helix     & $76.4\pm0.7$ & ($23.8$\ldots$22.7$)$\pm0.4$  & 23.33& (3,2) \\[-.5pt]
(4,1) &	0.78561&  77.4148& 0.765\ldots0.808 & 5-helix$^\ast$     & $77.5\pm0.7$ & ($11.0$\ldots$10.2$)$\pm0.6$  & 10.72& (4,1) \\[-.5pt]
(5,0) &	0.85065&  72.0000& 0.819\ldots0.851 & (5,0)       & $72.0\pm0.7$ & $0.0\pm0.6$    &  0.00& (5,0) \\[-.5pt]
(3,3) &	0.86603&  60.0000& 0.861            & (3,3)       & $60.0\pm0.5$ & $29.9\pm0.4$                    & 30.00& (3,3) \\[-.5pt]
(4,2) &	0.88462&  64.6055& 0.872\ldots0.904 & 6-helix     & $64.7\pm0.9$ & ($19.2$\ldots$18.3$)$\pm0.5$  & 19.01& (4,2) \\[-.5pt]
(5,1) &	0.93259&  63.9796& 0.914\ldots0.957 & 6-helix     & $64.0\pm0.8$ & ($9.0$\ldots$8.4$)$\pm0.6$    &  8.84& (5,1) \\[-.5pt]
(6,0) &	1.00000&  60.0000& 0.967\ldots      & (6,0)       & $60.0\pm0.6$ & $0.0\pm0.5$                     &  0.00& (6,0) \\[-.5pt]
(4,3) &	1.00188&  53.6574& \ldots1.021      & 7-helix     & $53.8\pm0.6$ & $26.2\pm0.7$ & 25.26& (4,3) \\[-.5pt]
(5,2) &	1.03116&  55.5587& 1.031\ldots1.052 & 7-helix     & $55.5\pm0.6$ & ($15.9$\ldots$15.5$)$\pm0.6$  & 16.02& (5,2) \\[-.5pt]
(6,1) &	1.08319&  54.4683& 1.063\ldots1.106 & 7-helix     & $54.5\pm0.6$ & ($7.6$\ldots$7.2$)$\pm0.5$    &  7.52& (6,1) \\[-.5pt]
(4,4) &	1.13152&  45.0000& 1.106\ldots1.127 & (4,4)       & $45.0\pm0.5$ & ($30.3$\ldots$30.0$)$\pm0.5$  & 30.00& (4,4) \\[-.5pt]
(5,3) &	1.14441&  47.8827& 1.138            & 8-helix     & $47.9\pm0.3$ & $21.7\pm0.4$                    & 21.75& (5,3) \\[-.5pt]
(7,0) &	1.15238&  51.4286& 1.148\ldots1.169 & (7,0)       & $51.4\pm0.4$ & $0.0\pm0.6$ &  0.00& (7,0) \\[-.5pt]
(6,2) &	1.18076&  48.5578& 1.169\ldots1.201 & 8-helix     & $48.6\pm0.5$ & ($13.9$\ldots$13.4$)$\pm0.5$  & 13.83& (6,2) \\[-.5pt]
(7,1) &	1.23600&  47.3936& 1.212\ldots1.254 & 8-helix     & $47.4\pm0.4$ & ($6.7$\ldots$6.4$)$\pm0.5$    &  6.54& (7,1) \\[-.5pt]
(5,4) &	1.26887&  41.3657& 1.244\ldots1.276 & 9-helix     & $41.3\pm0.3$ & ($26.8$\ldots$26.0$)$\pm0.6$                    & 26.32& (5,4) \\[-.5pt]
(6,3) &	1.29090&  42.9481& 1.286            & 9-helix     & $42.9\pm0.3$ & $19.0\pm0.5$  & 19.07& (6,3) \\[-.5pt]
(8,0) &	1.30656&  45.0000& 1.297\ldots1.318 & (8,0)       & $45.0\pm0.4$ & $0.0\pm0.6$                     &  0.00& (8,0) \\[-.5pt]
(7,2) &	1.33242&  43.0405& 1.318\ldots1.361 & 9-helix     & $43.0\pm0.5$ & ($12.2$\ldots$11.8$)$\pm0.6$  & 12.17& (7,2) \\[-.5pt]
(8,1) &	1.39027&  41.9316& 1.371\ldots1.424 & 9-helix     & $41.9\pm0.5$ & ($5.8$\ldots$5.6$)$\pm0.8$    &  5.79& (8,1)
\end{tabular*}
}}
\caption{Characteristics of lowest-energy states of simulated extendible polymers on cylindrical surfaces with radius $r$. The given angles are, as before, average values based on local measurements. See text for a detailed description. Structure types marked with an ``$^\ast$'' are visualized in Fig.~\ref{fig:7-tube_helices}.}
\label{tab:2}
\end{table}

Simulations were now performed for different $r$ values in the range
$r\in[0.425,\ldots,1.424]$ independently, with a step width of
about~$0.01$.  The lengths of the simulated polymer
chains ranged from $N=32$ to $200$. The results presented below
also did not depend on the actual number of monomers~$N$.

As mentioned above, we are looking for ground states in order to find
a classification scheme for polymer nanotubes that depends on their
radius, analogously to that of the carbon nanotubes. Indeed, this can
be reduced to the measurement of the characteristic angles $\psi$ and
$\theta$ of such lowest-energy polymer tubes, which we measure as
described above, depending on the given radius. For defect-free
conformations, this can then be related to known observables of
SWCNTs.

\begin{figure}[b!]
\begin{center}
\includegraphics[width=.95\textwidth,clip]{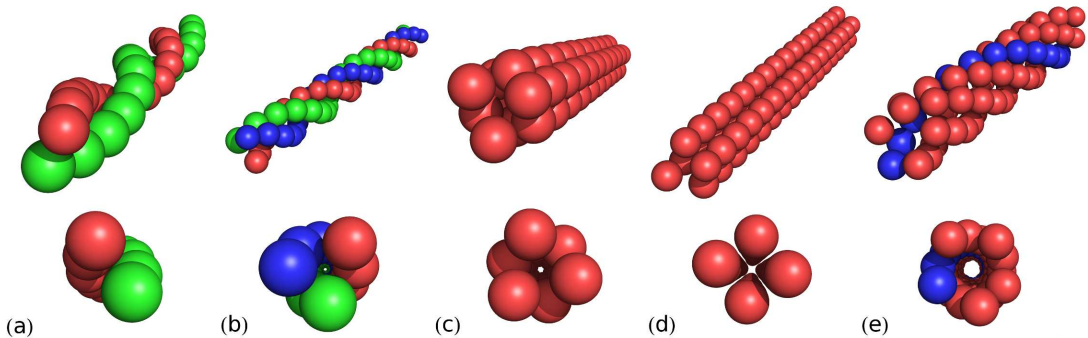}
\end{center}
\caption{Putative ground-state structures of the 
  polymer model restricted to a cylinder surface.  The
  structures were found at different tube radii and belong to
  different chirality classes. (a) Double-helix; (b) Triple-helix, (2,1)
  tube; (c) (3,0) tube; (d) (2,2) tube; (e) 5-helix, (4,1) tube. Every
  picture shows a single polymer, whereas just the monomer positions
  are shown and not the bonds between consecutive monomers. Different
  colors were used to mark imaginary helical strands. The top row
  shows a perspective view on the structures, the bottom row a top
  view.}
\label{fig:7-tube_helices}
\end{figure}

Table~\ref{tab:2} summarizes results from this part of our
computational study, and in Fig.~\ref{fig:7-tube_helices} we visualize
some of the putative ground-state structures we found that belong to different
chirality classes. The illustrated structures are marked in
Table~\ref{tab:2} with an asterisk. The first column shows an $(n,m)$
tube code, the second and third column the corresponding calculated
radius \smash{$r_{\mathrm{poly},\vartriangle}^{(n,m)}$} and the angle
$\psi$ in the polyhedral model for triangular nanotubes (cf.
Eqs.~(\ref{eq:polyrad}) and~(\ref{eq:psi})). The rows are ordered with
respect to increasing values of
\smash{$r_{\mathrm{poly},\vartriangle}^{(n,m)}$}. In the fourth
column, the input radius is given. Since transitions between different
chiralities are not continuous, we give an interval of radii, for
which the lowest-energy structures of the polymer fall into the same
chirality class, which is given in the fifth column. By ($n+m$)-helix
we denote structures which can be considered to be composed of $n+m$
virtual, interwoven chains of monomers with a helical wrapping, see
Fig.~\ref{fig:7-tube_helices} for further clarification.  The average
values of the angles $\psi$ and $\theta$ measured in the lowest-energy
conformations are listed next to them.  In the last two columns, known
wrapping angles from \hbox{SWCNTs} could be uniquely assigned to the
results from our SWPT simulations along with the corresponding SWCNT
types.

\subsection{Reproducing SWCNT sequences of chirality}
\label{sec:OA}

If one sorts SWCNTs with respect to their radii, a specific sequence
of chiralities is found. This sequence is exactly reproduced by
polymer monolayer tubes, i.e., we can confirm the assumed correlation
between SWCNTs and SWPTs~\cite{vb1}. In Table~\ref{tab:2},
we also compare the radii of ($n,m$) polymer tubes inserted into the
simulations (column~4) with the exact values obtained from
Eq.~(\ref{eq:polyrad}) (column~2), as well as with the characteristic
angles $\psi$ and $\theta$ (columns 3~vs.~6 and
7~vs.~8). Our simulation results agree
perfectly with the predictions from the respective polyhedral
model~\cite{lee3} and the high accuracy allow for the identification
of the chiralities of the polymer tubes.  Nonetheless, there are
certain regions of radii, where the difference of radii between
different tube types is extremely small. Within these regions, it is
particularly challenging to resolve explicitly different chiralities
(see the ``plateaus'' in Fig.~5 in Ref.~\cite{vb_athens11}).  In
Table~\ref{tab:2}, results for (6,0) and (4,3) tubes are therefore
listed in the same row. Their radii in the polyhedral model differs by
less than $0.2\,\%$, which is reflected accordingly in our results.
Regarding the wrapping angles, we could hence reproduce exactly the
characteristic sequence for SWCNTs. Together with the considerations
about the scaling of lengths between SWCNTs (cf. Fig.~\ref{fig:5})
this is the link between ideal SWCNTs and \hbox{SWPTs}.

\begin{figure}[b!]
\begin{center}
\includegraphics[width=.7\textwidth,clip]{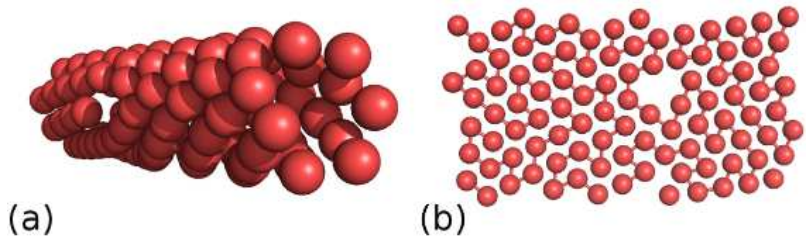}
\end{center}
\caption{Low energy state with a defect and two regions with different
  wrappings. The radius of the tube is $r_\mathrm{input}\approx1.11$,
  the competing substructures correspond to (6,1)- and (5,3)-tubes. In
  (b) the conformation is shown in an unzipped view for clarity.}
\label{fig:8}
\end{figure}

We would like to emphasize that this accuracy is essential as one
would not have been able to draw these detailed conclusions using the
conventional approach (Eq.~(\ref{eq:convrad})) for the radius
calculation. The polymer model with slightly different parameters that
was used in Ref.~\cite{vb1} already yielded the correct trends
of the present results, but only the flexible bond-length model
allowed us to study the details precisely and quantitatively correct.
In this case, we find transitions at the interfaces between two
structural regions. We observe in the ensemble of low-energy states
``competing'' conformations, i.e., different tube types with very
similar ground state energies and tubes with defects or internal
interfaces between regions belonging to different chirality classes,
see Fig.~\ref{fig:8} for\break{} examples.

\subsection{The link to the internal structure of 'phase B'}

Actually, the intriguing monolayer polymer structures we found in
our previous study~\cite{vb1} for strong wire adsorption, and which we
summarized under the name 'barrel (B) phase', made us think about a
possible link to single-walled carbon nanotubes and were the
motivation for the present study. Beside the systematic presentation of
the results above, let us therefore comment on some actual structures
we found earlier.

\begin{figure}[b!]
\begin{center}
\includegraphics[width=.7\textwidth,clip]{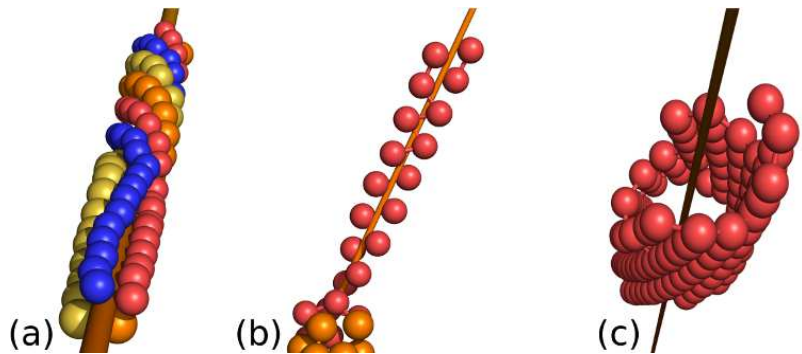}
\end{center}
\caption{Monolayer polymer tubes in the barrel phase B for strong wire
  adsorption. (a) $\sigma_\mathrm{f}=0.65$, (b)
  $\sigma_\mathrm{f}=0.40$, and (c) $\sigma_\mathrm{f}=1.57$.
  Different colors or shadings shall facilitate the perception only,
  data taken from study presented in Ref.~\cite{vb1}.}
\label{fig:9}
\end{figure}

At $\sigma_\mathrm{f}=0.647$, we found in a monolayer
polymer conformation consisting of two competing substructures with
different chirality and a defect at the interface between both
substructures (see  Fig.~\ref{fig:9}\,(a))~\cite{vb1}. One region forms a
$(2,2)$ tube, the other is a 4-helix with a measured mean wrapping
angle of $\theta=14\pm4$, which corresponds to a~$(3,1)$ tube. These
observations fit in perfectly with the results presented above in
Table~\ref{tab:2}. Both structures are neighbors in the radius-ordered
sequence of tube structures. Slight deviations of the value of the
radius and the larger error of the mean wrapping angle trace back to
the fact that we originally used a slightly different polymer model
with fixed bond length (sticks instead of elastic bonds) where the
optimal distance between two nonbonded monomers was slightly larger
than the bond length.  However, on the other hand, this indicates
that our results are of general character and do not depend on
certain details of the implemented polymer model.
Another conformation found in the earlier study was the, somewhat
artificial, (1,1) tube at $\sigma_\mathrm{f}=0.4$ (see
Fig.~\ref{fig:9}\,(b)). Again, this was confirmed exactly by the
present study and fits into the general scheme as presented here.

Finally, let us look at conformations with larger radius of
$r\approx1.7$, as shown in Fig.~\ref{fig:9}\,(c). There, we found a
12-helix with chiral angle of $20\pm1^\circ$, which can now be
assigned to a (8,4) tube (not listed in Table~\ref{tab:2}, but
calculations lead to $\theta^{(8,4)}=19.1$ and
$r_{\mathrm{poly},\vartriangle}^{(8,4)}=1.71$). However, we also found
in that region a strong competition between conformations with
different structures as well as conformations composed of different\vadjust{\break}
substructures, as shown for example in Figs.~1 and~4 in
Ref.~\cite{vb_athens10}. We find for example regions which can
be explained to be parts of (7,5) tubes (not in Table~\ref{tab:2},
$\theta^{(7,5)}=24.5$, $r_{\mathrm{poly},\vartriangle}^{(7,5)}=1.68$).
This structure forms also a 12-helix, but with just slightly different
wrapping angle compared to (8,4). Finally, we find among
those structures (10,1) tubes ($\theta^{(10,1)}=4.7$,
$r_{\mathrm{poly},\vartriangle}^{(10,1)}=1.70$ forming 11-helices.
Note the small differences between
$r_{\mathrm{poly},\vartriangle}^{(8,4)}$,
$r_{\mathrm{poly},\vartriangle}^{(10,1)}$, and
$r_{\mathrm{poly},\vartriangle}^{(7,5)}$.

\section{Summary}

In this study, we have investigated the relationship between
single-walled carbon nanotubes and the tube phase of a bead--spring
polymer model attracted by a thin wire. In fact, we found surprisingly
clear geometrical similarities between these different materials. We
could only obtain our results by taking into account the discreteness
and curvature effects in the mathematical description of the
geometrical properties of nanotubes.  Hence, we provide an example for
the necessity of applying accurate discrete models, rather than
continuous approximations, in computational studies of nanotubes.

To strengthen our theoretical considerations on the link between
carbon nanotubes and polymer nanotubes, we employed numerical
optimization procedures to construct lowest-energy polymer
conformations for given attraction length scales of the wire (or,
equivalently, given polymer tube radii). Comparing those conformations
based on a triangular lattice, with carbon nanotubes based on a
hexagonal honeycomb lattice, we found that both share the same
chirality sequence and we show how the length scales are
connected. We addressed the problem of competing substructures leading
to defects in non-ideal structures, which definitely merits further
investigation.

The perfect structural coincidence between atomic nanotubes and
polymer tubes explains the internal structure of the barrel phase of
polymers adsorbed at nanowires. It is also a good starting point for
the further systematic investigation of hybrid systems of polymers and
single-walled carbon nanotubes, which might be of technological
interest for controlling, e.g., physical properties of polymer-coated
nanotubes.  We have also presented further evidence, that helical
conformations are intrinsic natural structures in simple polymer
models~\cite{vogel09epje}.

Our key result of the universal nature of the sequence of
conformations in nanotubes and polymer tubes has developed from a
series of simulations.  Such extensive 3d simulations have only
recently become possible, and the situation is reminiscent of the
discoveries in the seventies and early eighties as the picture of
universality in critical phenomena emerged from the early numerical
results of series expansions and simulations, and a very few
experiments.  With current and future computer power, the two
directions mentioned above, namely the exploration of defective
structures and the study of hybrid systems should lead into
interesting theoretical and practical directions in polymer research
and nanotechnology.

\section*{Acknowledgments}
  The authors would like to thank P.~Pine and S.~Srebnik from the
  Technion Haifa, for valuable discussions on nanotubes and adsorption
  of polymers at nanotubes. This project was in part supported by the
  J\"ulich/Aachen/Haifa Umbrella program under Grants No.~SIM6 and
  No.~HPC\_2. Supercomputer time was provided by the Forschungszentrum
  J\"ulich under Projects No.~jiff39 and No.~jiff43.


\begin{thebibliography}{99}
\bibitem{footnote1} In analogy to the length $l_\mathrm{CC}$
  introduced for the C--C bond length in SWCNTs. For $n\to\infty$,
  $\sqrt{3}\,l_\mathrm{CC}=r_0$ (cf.~also Eq.~(\ref{eq:Lconv})) for
  tubes with equal radii.  The scaling between the lengths in
  triangular and honeycomb tubes for finite $n$ will be discussed
  below.
\end{thebibliography}
\end{document}